**Nonlinear dynamics and bifurcations of a planar undulating magnetic microswimmer**

Jithu Paul, Yizhar Or*, Oleg V. Gendelman

Faculty of Mechanical Engineering, Technion – Israel Institute of Technology, Technion City, Haifa 3200003, Israel

* - contacting author, e-mail: izi@me.technion.ac.il

**Abstract**

Swimming micro-organisms such as flagellated bacteria and sperm cells have fascinating locomotion capabilities. Inspired by their natural motion, there is an ongoing effort to develop artificial robotic nano-swimmers for potential in-body biomedical applications. A leading method for actuation of nano-swimmers is by applying a time-varying external magnetic field. Such systems have rich and nonlinear dynamics that calls for simple fundamental models. A previous work studied forward motion of a simple two-link model with passive elastic joint, assuming small-amplitude planar oscillations of the magnetic field about a constant direction. In this work, we found that there exists a faster, backward motion of the swimmer with very rich dynamics. By relaxing the small-amplitude assumption, we analyze the multiplicity of periodic solutions, as well as their bifurcations, symmetry breaking, and stability transitions. We have also found that the net displacement and/or mean swimming speed are maximized for optimal choices of various parameters. Asymptotic calculations are performed for the bifurcation condition and the swimmer's mean speed. The results may enable significantly improving the design aspects of magnetically-actuated robotic microswimmer.

## 1 Introduction

The fascinating locomotion capabilities of swimming micro-organisms has attracted attention of scientific community for long time. Owing to the small scale of such swimmers, their motion is governed by low Reynolds number hydrodynamics, where viscous drag forces are dominating while inertial effects are negligible [1]. Several mathematical models of undulatory micro-swimming have been studied, dating back to classic work of Taylor on infinite wavy sheet [2] and later works of Lighthill [3] and Childress [4]. The well-known work by Purcell [5] introduced simplified robotic-like models inspired by swimming microorganisms, the corkscrew motion of counter-rotating spherical head and helical tail, as well as planar motion of three-link model with controlled joint angles whose motion has been further analyzed in many follow-up works [6–8]. Another important effect that has been suggested in [5] is combining time-periodic actuation with body's flexural elasticity, which has also been studied in analytical and numerical models of sperm motility [9,10].

In the last two decades, observations from natural swimming microorganisms have inspired the ongoing development of engineered artificial nano-micro-scale swimmers, aiming towards in-body biomedical applications such as targeted drug delivery, diagnosis and minimally-invasive operation [11]. Few actuation mechanisms have been considered for powering such nano-swimmers, including chemical activation [12] as well as bio-hybrid swimmers harnessing the actual beating of live bacterial flagellum [13]. Nevertheless, a leading concept for nano-swimmers actuation is using time-varying external magnetic field. While several works have realized corkscrew locomotion induced by rotating magnetic field [14], the pioneering work of Dreyfus et al [15] actually used planar oscillating magnetic field for propelling a chain of superparamagnetic beads connected by flexible DNA link to a "head" made of red blood cell. The magnetic field in [15] was set to be spatially uniform and time-varying, as

$$\mathbf{B}(t) = c\hat{\mathbf{x}} + b\sin(\Omega \tau)\hat{\mathbf{y}} \tag{1}$$

where $b, c \geq 0$ are constants. Thanks to ongoing progress in nano-fabrication capabilities, simpler designs of nano-swimmers composed of rigid links connected by flexible hinges were later proposed [16,17]. A simple theoretical model for studying the planar locomotion of such swimmer is the two-link model proposed in [18], see Fig. 1. This model consists of two rigid links connected by a passive elastic joint represented as a torsion spring, and one of the links (the "head") is magnetized along its longitudinal axis. The analysis in [18] focused on the case of small oscillations $b \ll c$ and conducted asymptotic analysis of the motion in which the swimmer oscillates about and swims along $+\hat{\mathbf{x}}$ direction, which is a stable periodic solution with mean orientation angle $\bar{\theta} = 0$. The analysis showed that there exist optimal actuation frequencies $\Omega$ for maximizing the mean speed or displacement per cycle. In this work, we revisit the two-link model in [18] and extend the analysis to cases of large oscillations $b > c$ and even $c = 0$, and study also the "backward" solution where the swimmer oscillates about and swims along $-\hat{\mathbf{x}}$ direction, with $\bar{\theta} = \pi$. While this swimmer's orientation $\theta = \pi$ is statically unstable (for $b = 0, c > 0$), we find that for $b \neq 0$, this gives a periodic solution which undergoes stability transition and subcritical pitchfork bifurcation upon varying amplitude $b$ and frequency $\Omega$ of the magnetic field's input. We analyze the backward solution numerically as well as analytically using asymptotic expansion and harmonic balance. Under small-angle expansion, the system's dynamics can be reduced to a nonlinear second-order differential equation with parametric excitation, which resembles the well-known Kapitza pendulum system [19,20]. Finally, we show optimization of the swimmer's net motion with respect to both $b$ and $\Omega$. Remarkably, we find that the optimal "backward" motion is faster than the forward motion.

## 2   Problem statement

The model consists of two rigid links representing head and tail, connected by a passive torsional spring with a linear stiffness $k$ (see Fig. 1). $x, y$ are position of head link's center point. The head makes an angle $\phi$ with the tail and an angle $\theta$ with the $\hat{\mathbf{x}}$-axis. The head is magnetized with a magnetization strength $h$ along its longitudinal axis t (bold hat), given by $\hat{\mathbf{t}}$ is the unit vector along the head's longitudinal direction given by $\hat{\mathbf{t}} = \cos\theta\hat{\mathbf{x}} + \sin\theta\hat{\mathbf{y}}$. The microswimmer is submerged in a Newtonian fluid and subjected to an external magnetic field. The magnetic field is spatially uniform, and has a time varying term in the $\hat{\mathbf{y}}$ direction and may or may not have a constant term in the $\hat{\mathbf{x}}$ direction. The magnetic field is represented as $\mathbf{B}(t) = c\hat{\mathbf{x}} + b\sin(\Omega t)\hat{\mathbf{y}}$ (see Fig. 1). Here, $c$ is the zero or nonzero constant in the $\hat{\mathbf{x}}$ direction, $b$ is the amplitude of oscillation in the $\hat{\mathbf{y}}$ direction and $t$ is time. The torque applied by the magnetic field on the swimmer's head link is given by $\mathbf{L} = h\hat{\mathbf{t}} \times \mathbf{B}(t)$.

The micro size of the swimmer allows making the following assumptions. For the swimmer, gravity is neglected and it is neutrally buoyant. As for the hydrodynamics, viscous forces much dominate over inertial forces and allow to assume low Reynolds number and neglecting inertial effects in the dynamics. Stokes' law governs the fluid motion and resistive force theory (RFT) [18,21,22] gives the final dynamics of the system. The viscous drag force $\mathbf{f}_i$ and torque $m_i$ under planar motion is considered as proportional to its linear ($\mathbf{v}_i$) and angular velocities ($\omega_i$) as follows.

$$\mathbf{f}_i = -c_t l(\mathbf{v}_i \cdot \hat{\mathbf{t}}_i)\hat{\mathbf{t}}_i - c_n l(\mathbf{v}_i \cdot \hat{\mathbf{n}}_i)\hat{\mathbf{n}}_i; \quad m_i = -\frac{1}{12}c_n l^3 \omega_i; \quad c_n = 2c_t \tag{2}$$

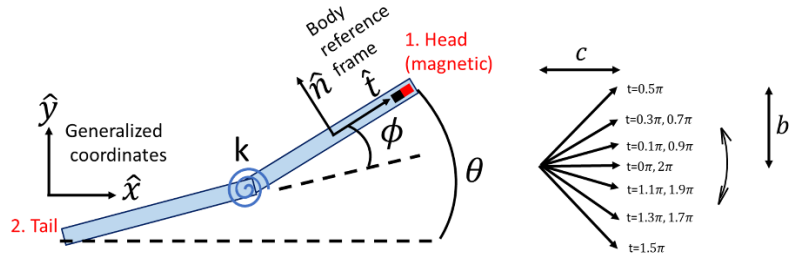

Fig. 1. The two-link microswimmer model [18].

In (2), $l$ is the length of the link, $\hat{\mathbf{t}}_i$ and $\hat{\mathbf{n}}_i$ are the unit vectors in the axial and normal directions of the i[th] link. The internal torque applied at the joint by the torsional spring is given by $\tau = -k\phi$.

The net force and torque balance gives (see Appendix A) the following final set of equations (Eq. (3)) for the four degrees of freedom of the swimmer. In (3), $v_t$ and $v_n$ are the tangential and normal velocities with respect to the head and can be converted into the world's frame by using a rotation matrix, as given in the appendix.

$$\begin{pmatrix} v_t \\ v_n \\ \dot{\theta} \\ \dot{\phi} \end{pmatrix} = \begin{pmatrix} -\dfrac{3\sin(\phi)(\cos(\phi)(bh\sin(\theta)+k\phi-bh\cos(\theta)\sin(t\Omega))+3k\phi)}{c_t l^2 \left(\cos^2(\phi)-9\right)} \\ \dfrac{6\left(\cos^2(\phi)(c\sin(\theta)+k\phi-bh\cos(\theta)\sin(t\Omega))-\left(\sin^2(\phi)+5\right)(ch\sin(\theta)+k\phi-bh\cos(\theta)\sin(t\omega))+4k\phi\cos(\phi)\right)}{c_t l^2 \left(68-4\cos(2\phi)\right)} \\ \dfrac{3\cos^2(\phi)(ch\sin(\theta)+k\phi-bh\cos(\theta)\sin(t\Omega))-3\left(\sin^2(\phi)-19\right)(ch\sin(\theta)+kh\phi-bh\cos(\theta)\sin(t\Omega))+36k\phi\cos(\phi)}{c_t l^3 \left(\cos(2\phi)-17\right)} \\ \dfrac{6(\cos(\phi)+3)^2(ch\sin(\theta)+2k\phi-bh\cos(\theta)\sin(t\Omega))}{c_t l^3 \left(\cos(2\phi)-17\right)} \end{pmatrix}$$

(3)

The system of equations (3) can be non-dimensionalized by defining $\gamma = \dfrac{ch}{k}$, $\beta = \dfrac{bh}{k}$ and $t_k = \dfrac{c_t l^3}{k}$ (the elastic time scale). In addition, we rescale the time in (3) as $t \to \dfrac{t}{t_k}$ and rescale the frequency by $\omega = \Omega t_k$. Then (3) will be simplified and nondimensionalized as (4), for $\dot{\theta}$ and $\dot{\phi}$,

$$\begin{pmatrix}\dot{\theta}\\ \dot{\phi}\end{pmatrix}=\begin{pmatrix}\dfrac{3\cos^2(\phi)(\gamma\sin(\theta)+\phi-\beta\cos(\theta)\sin(t\omega))-}{\cos(2\phi)-17}\\ \dfrac{-3(\sin^2(\phi)-19)(\gamma\sin(\theta)+\phi-\beta\cos(\theta)\sin(t\omega))+36\alpha\phi\cos(\phi)}{\cos(2\phi)-17}\\ \dfrac{6(\cos(\phi)+3)^2(\gamma\sin(\theta)+2\phi-\beta\cos(\theta)\sin(t\omega))}{\cos(2\phi)-17}\end{pmatrix} \qquad (4)$$

Due to positional symmetry of Eq. (3), Eq. (4) determines the entire dynamics of the system. When $\theta(t)$ and $\phi(t)$ are solved, $v_t$ and $v_n$ can be immediately calculated from Eq. (3). The system's solution thus depends on three dimensionless parameters, $\gamma, \beta$ and $\omega$, and on initial conditions.

In the following sections we will integrate (3) or (4) numerically using MATLAB packages, and analytically by using perturbation expansion and harmonic balance [23,24].

## 3  Results

### 3.1  Numerical treatment

The numerical strategy to solve the set of equations using MATLAB is described here. The steady state solutions of the swimmer are periodic in $\theta, \phi$ and the dynamics are invariant with respect to $x, y$ and so we define a reduced state vector $\mathbf{z}(t)=(\theta(t),\phi(t))^T$. We also denote a function $\mathbf{F}(\mathbf{z}(0))=\mathbf{z}(T)$, where $\mathbf{z}(0)$ the system's is initial condition and $T=\dfrac{2\pi}{\omega}$ is the nondimensional time period. This function is also known as the stroboscopic map [25] (we sample $\mathbf{z}(t)$ in fixed rate). System's periodic solutions correspond to initial conditions that satisfy $\mathbf{z}^*=\mathbf{F}(\mathbf{z}^*)$. By solving the equation using MATLAB's solver of nonlinear equations *fsolve*, we obtain fixed points of $\mathbf{F}$, which correspond to periodic solutions of the system. The stability of the periodic solution is determined by calculating the Jacobian matrix of $\mathbf{F}$: $\mathbf{J}=\dfrac{d\mathbf{F}}{d\mathbf{z}}$ at $\mathbf{z}=\mathbf{z}^*$. Calculating the eigen values $\lambda_i$ of $\mathbf{J}$, the condition for asymptotic stability of the periodic solution is given by $|\lambda_i(\mathbf{J})|<1$.

In Fig. 2(a) and 2(b), we plot representative solution trajectories in the plane of two angles $\theta$ and $\phi$, which are the degrees of freedom in the reduced system of Eq. (4). As clear from the plots, co-existing symmetric periodic solutions with mean values $(\bar{\theta},\bar{\phi})=(0,0)$ [18] and $(\bar{\theta},\bar{\phi})=(\pi,0)$, are observed, representing forward and backward motion, respectively. However, in Fig. 2(a), backward motion is not stable whereas in Fig. 2 (b), it is stable, depending on different values of $\omega$ and $\beta$. In case of 2(b), each of the two stable periodic solution has its own basin of attraction - region of initial conditions $(\theta(0),\phi(0))$ that converge to it. A grid discretization of these regions, colored in blue and red, is shown in Fig. 2(c). Note that the angle theta is $2\pi$-periodic. The difference between the cases in Fig. 2(a) and 2(b) clearly indicates that unlike the forward motion which is always stable [18], there exists stability transitions in the backward motion, and those stability transitions must be accompanied by bifurcations. The stability transition curves, obtained numerically for different parameter ranges, are given in Fig. 3. When the backward solution with $(\bar{\theta},\bar{\phi})=(\pi,0)$ undergoes transition from unstable to stable, a pair of asymmetric branches of unstable periodic solutions begins to evolve (Fig. 4), where $(\bar{\theta},\bar{\phi})\neq(\pi,0)$. These unstable asymmetric solutions are denoted as purple dashed loops in Fig. 2(b). As plotted in Fig. 4, we captured subcritical pitchfork bifurcation at this stability transition. In addition, it is interesting to note that, there exist optimum mean speed $V=\dfrac{x(T)-x(0)}{T}$ and net displacement $X=x(T)-x(0)$ with respect to $\beta$ and $\omega$ and the optimum values can be tuned into stable region by setting $\gamma \to 0$ for realistic parameter values, as shown in Fig. 5.

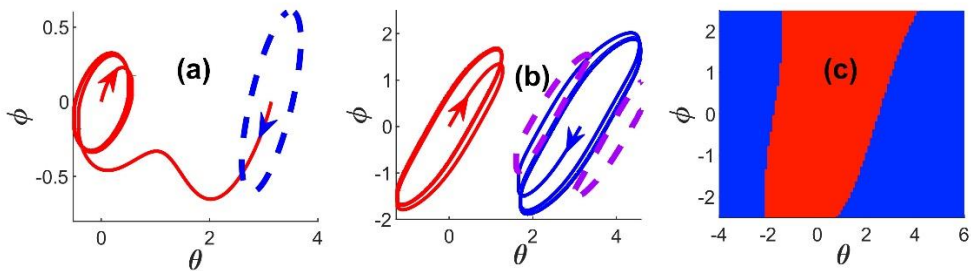

Fig. 2. Solution trajectories in $(\theta,\phi)$ plane. Stable periodic solutions oscillating about $\bar{\theta}=0$ and $\bar{\theta}=\pi$. In (a), $\omega=2.5, \gamma=1$ and $\beta=1.1$. Red solid line shows forward stable solution oscillating about mean values $(\bar{\theta},\bar{\phi})=(0,0)$ and blue dashed curve shows unstable backward solution oscillating about mean values $(\bar{\theta},\bar{\phi})=(\pi,0)$. In (b), $\omega=10, \gamma=1$ and $\beta=10$. Red solid line shows stable solution around $(\bar{\theta},\bar{\phi})=(0,0)$, blue solid line shows stable solution around $(\bar{\theta},\bar{\phi})=(\pi,0)$ and purple dashed lines show asymmetric unstable periodic solutions. In (c), a grid discretization is shown for the regions of attractions of initial conditions that converge to either the forward (red) or backward (blue) periodic solution of the plot (b).

Remarkably, Fig. 5 also shows that the swimmer goes faster in the backward direction compared to forward direction, for $\gamma \neq 0$. This effect is even amplified for larger $\beta$ and $\gamma$, see Fig. 5(b).

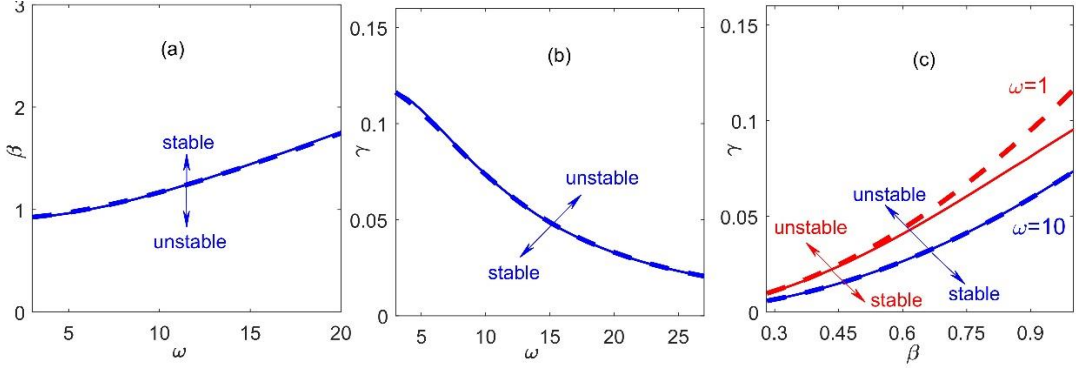

Fig. 3. Stability transition curves of the symmetric backward periodic solution with $(\bar{\theta},\bar{\phi})=(\pi,0)$. Solid lines show the numerical result and dashed lines show the asymptotic calculation obtained below in in Eq. (19). In (a) $\gamma = 0.1$, and in (b) $\beta = 1$.

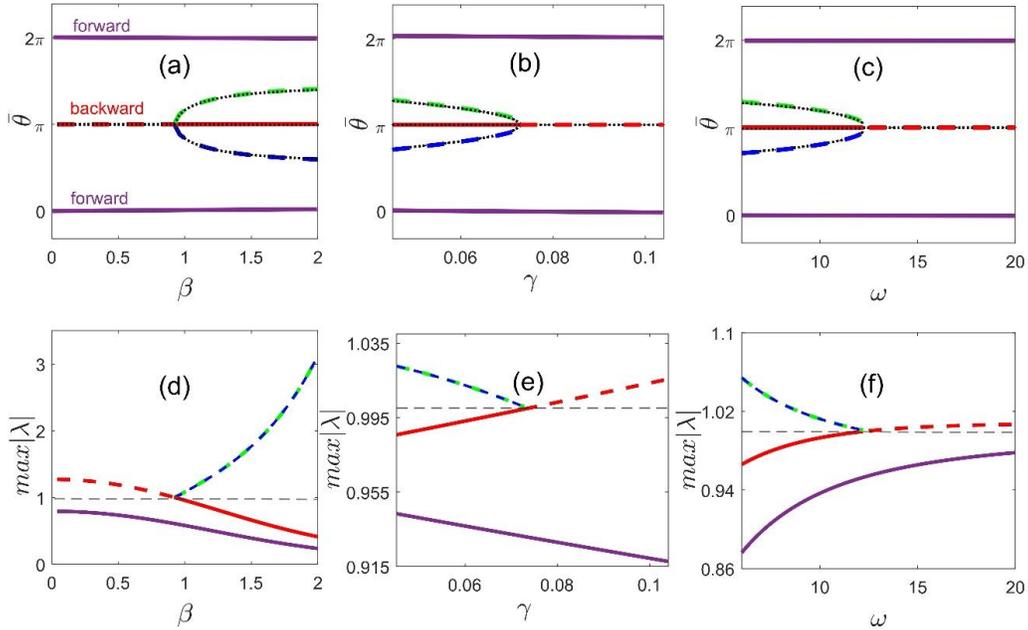

Fig. 4. In (a), (b), (c), stability transitions of the backward periodic solution with $\bar{\theta} = \pi$, accompanied with subcritical pitchfork bifurcation in $\bar{\theta}$ versus $\beta, \gamma, \omega$, respectively, are shown. The color lines show results from numerical calculation. The red dashed and solid lines show symmetric solutions. The green and blue dashed lines shows asymmetric solution. The purple solid lines show the forward solution $\bar{\theta} = 0$. Note that $\theta$ is $2\pi$-periodic, so that $\bar{\theta} = \{0, 2\pi\}$ denote the same solution. Black dotted curves shows analytical calculations from harmonic balance (see section 3.2.1). In (d), (e), (f), the maximal eigenvalue max $|\lambda|$ with variation of parameters is shown, where crossing $|\lambda| = 1$ indicates stability transition. Here, in (a) and (c), $\omega = 2, \gamma = 0.1$; in (b) and (e), $\beta = 1, \omega = 10$ and; in (c) and (f), $\beta = 1, \gamma = 0.06$.

In addition, there is nonzero net propulsion in the case of $\gamma = 0$, i.e. zero mean of the field $\mathbf{B}(t)$, which has not been considered in [18].

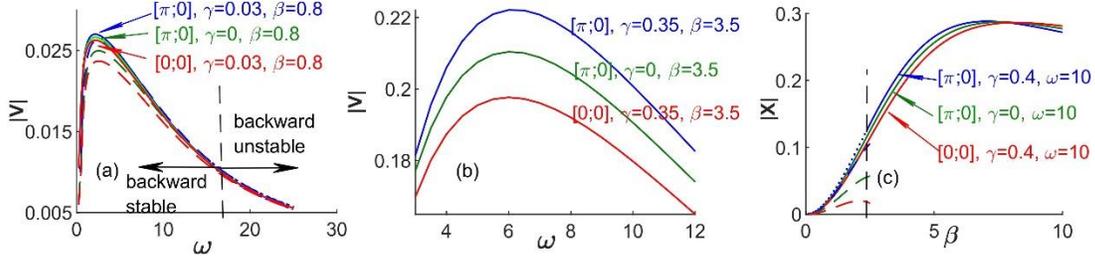

Fig. 5. In (a) and (b) The mean speed $V$ versus frequency $\omega$, showing existence of optimum. Solid lines show numerical calculation and dashed lines show analytical calculation using Eq. (23). (c) shows numerical calculation of net displacement X versus amplitude of actuation oscillations $\beta$, also indicating an optimal value. Here, the analytical calculation (dashed curves) diverges with $\beta$. The vertical black lines indicate stability transition of the backward solution and the dotted curve shows the unstable region. Red line shows $(\bar{\theta},\bar{\phi})=(0,0)$ branch with nonzero $\gamma$, blue line shows $(\bar{\theta},\bar{\phi})=(\pi,0)$ branch with nonzero $\gamma$ and green line shows $(\bar{\theta},\bar{\phi})=(0,0)$ branch with $\gamma=0$.

### 3.2 Analytical Investigation

The approach we adopt here is a combination of two methods: first, small-angle linear approximation in the joint angle $\phi$, results in a second-order nonlinear equation in $\theta$ only, which accurately captures (numerically) the dynamic. Next, we also expand to second order in $\theta$ about $\pi$ and followed by harmonic balance approximation.

Expanding all terms in $\phi$ about $\phi=0$ in Eq. (4) and taking only up to linear terms gives,

$$\begin{pmatrix} \dot{\theta} \\ \dot{\phi} \end{pmatrix} = \begin{pmatrix} \dfrac{60(\gamma\sin(\theta) - \beta\cos(\theta)\sin(t\omega)) + 96\phi}{-16} \\ -6(\gamma\sin(\theta) + 2\phi - \beta\cos(\theta)\sin(t\omega)) \end{pmatrix} \qquad (5)$$

$\phi$ can be eliminated from (5) using the D operator ($Df(t) = \dfrac{df}{dt}$) as follows. From the second equation of (5) we obtain,

$$(D+12)\phi = -6\gamma\sin(\theta) + 6\beta\cos(\theta)\sin(t\omega) \qquad (6)$$

Multiplying the first equation of (5) by $(D+12)$ and substituting Eq. (6), one obtains

$$(D+12)\dot{\theta} = \dfrac{60(D+12)(\gamma\sin(\theta) - \beta\cos(\theta)\sin(t\omega)) + 96(D+12)\phi}{-16} \qquad (7)$$

Eliminating $\phi$ by substitution of (6) into (7) gives a second-order nonlinear time-periodic differential equation in $\theta$ only, as

$$\ddot{\theta} + \frac{15}{4}\dot{\theta}(\beta \sin\theta \sin(t\omega) + \gamma \cos\theta) + 9\gamma \sin\theta +$$
$$+12\dot{\theta} - \frac{15}{4}\beta\omega \cos\theta \cos(t\omega) - 9\beta \cos\theta \sin(t\omega) = 0 \qquad (8)$$

Note that Eq. (8) is analogous to Kapitza pendulum [20], in the following sense. For zero excitation $\beta = 0$ and $\gamma > 0$, the system has a stable equilibrium point at $\theta = 0$ and an unstable one at the inverted position $\theta = \pi$. For nonzero excitation $\beta > 0$, there is a periodic solution oscillating around $\bar{\theta} = 0$ which is always stable, while stability of the solution with $\bar{\theta} = \pi$ may transition depending on system parameter values. (More precisely, the system in (8) is analogous to Kapitza pendulum with inclined base excitation, see details in [19]).

Taylor series (2nd order) expansion of $\sin\theta$ and $\cos\theta$ in Eq. (8) about $\theta = \pi$ gives;

$$\ddot{\tilde{\theta}} + \frac{15}{4}\dot{\tilde{\theta}}\left(-\beta\tilde{\theta}\sin(t\omega) + \gamma\left(-1 + \frac{1}{2}\tilde{\theta}^2\right)\right) - 9\gamma\tilde{\theta} + 12\dot{\tilde{\theta}} -$$
$$- \frac{15}{4}\beta\omega\left(-1 + \frac{1}{2}\tilde{\theta}^2\right)\cos(t\omega) - 9\beta\left(-1 + \frac{1}{2}\tilde{\theta}^2\right)\sin(t\omega) = 0 \qquad (9)$$

Where $\tilde{\theta} = \theta - \pi$.

For harmonic balance, we assume a periodic solution (truncating after first harmonics)

$$\tilde{\theta}(t) = a_0 + a_1 \cos(t\omega) + b_1 \sin(t\omega)$$
$$\dot{\tilde{\theta}}(t) = -a_1\omega\sin(t\omega) + b_1\omega\cos(t\omega) \qquad (10)$$
$$\ddot{\tilde{\theta}}(t) = -\omega^2\left(a_1 \cos(t\omega) + b_1 \sin(t\omega)\right)$$

Substituting (10) into (9) and rearranging gives

$$M_0 + \sum_{1}^{k} M_k \sin(k\omega t) + N_k \cos(k\omega t) = 0 \qquad (11)$$

Equating the coefficients of each harmonic to zero gives a polynomial system of three equations in the unknowns $a_0, a_1, b_1$:

$$M_0 = a_0\left(-\frac{9}{2}\beta b_1 - 9\gamma\right) = 0$$

$$M_1 = \frac{15}{8}a_0^2 b_1\gamma\omega - \frac{15}{8}a_0^2\beta\omega + \frac{15}{32}a_1^2 b_1\gamma\omega - \frac{15}{32}a_1^2\beta\omega - \frac{9}{4}a_1\beta b_1 -$$

$$-9a_1\gamma - a_1\omega^2 + \frac{15}{32}b_1^3\gamma\omega - \frac{45}{32}\beta b_1^2\omega + 12b_1\omega - \frac{15}{4}b_1\gamma\omega + \frac{15}{4}\beta\omega = 0 \quad (12)$$

$$N_1 = -\frac{15}{8}a_0^2 a_1\gamma\omega - \frac{9}{2}a_0^2\beta - \frac{1}{32}15a_1^3\gamma\omega - \frac{9}{8}a_1^2\beta - \frac{15}{32}a_1 b_1^2\gamma\omega +$$

$$+\frac{15}{16}a_1\beta b_1\omega - 12a_1\omega + \frac{15}{4}a_1\gamma\omega - \frac{27}{8}\beta b_1^2 - 9b_1\gamma - b_1\omega^2 + 9\beta = 0$$

Eq. (12) can be seen as $M_0 = a_0 Q(b_1) = 0$. Two types of possible solutions may exist, symmetric with $a_0 = 0$, and asymmetric pair with $Q = 0$ and $a_0 \neq 0$. Assuming $a_0 \neq 0$, note that equations $M_1 = 0$ and $N_1 = 0$ in (12) only involve quadratic terms of $a_0^2$, so we obtain two asymmetric solutions with $\pm a_0$ and same $a_1, b_1$. Solutions of (12) give nice plot of the pitchfork bifurcation (see Fig. 4) and the result is compared with numerical simulation, where the mean values are $\bar{\theta} = \pi \pm a_0$.

### 3.2.1 Bifurcation conditions

For the bifurcation, we require coincidence of both solution types, so $a_0 = Q = M_1 = N_1 = 0$. Extra fourth equation, gives conditions on parameters ($\beta, \gamma, \omega$) in addition to solving $a_1, b_1$.

$$M_0 = Q = 0 \text{ gives } a_0 = 0; b_1 = -\frac{2\gamma}{\beta} \quad (13)$$

$M_1 = 0$ gives;

$$a_1 = \frac{2\left(\frac{\sqrt{1296\beta^4\gamma^2 - 2304\beta^4\gamma\omega^2 - 5760\beta^2\gamma^3\omega^2 + 450\beta^6\omega^2 + 1125\beta^4\gamma^2\omega^2 + 64\beta^4\omega^4 - 900\gamma^6\omega^2}}{\beta^2} - 36\gamma - 8\omega^2\right)}{15\left(\frac{2\gamma^2\omega}{\beta} + \beta\omega\right)} \quad (14)$$

And finally $N_1 = 0$ gives;

$$9\beta - \frac{15}{32}a_1^3\gamma\omega - \frac{9}{8}a_1^2\beta - 12a_1\omega - \frac{15}{32}a_1 b_1^2\gamma\omega + \\ + \frac{15}{16}a_1\beta b_1\omega + \frac{15}{4}a_1\gamma\omega - \frac{27}{8}\beta b_1^2 - 9b_1\gamma - b_1\omega^2 = 0 \tag{15}$$

Substituting the expressions for $a_1$ and $b_1$ from (13) and (14) into (15) gives bifurcation condition as an equation on $\beta, \gamma, \omega$. This equation is cumbersome, and in the following, we try to simplify it by approximation under some scaling assumptions.

### 3.2.2 Asymptotic behavior of bifurcation condition

In order to simplify the expression in (15) for the bifurcation condition, we now assume the limit of small magnetic field (relative to spring stiffness) and fast oscillations. That is, the nondimensional parameters satisfy $\gamma \ll \beta \ll 1 < \omega$. More concretely, choosing some $\varepsilon \ll 1$, we assume scaling as $\gamma \sim O(\varepsilon^2)$, $\beta \sim O(\varepsilon)$, and $\omega \sim O(\varepsilon^{-0.5})$. We keep only terms up to $O(\varepsilon)$ while higher orders of $\varepsilon$ are neglected.

Based on the above assumption, $M_1 = 0$ in Eq. (12), for the symmetric solution $a_0 = 0$, can be approximated up to $O(\varepsilon)$ as,

$$-\frac{15}{32}a_1^2\beta\omega - a_1\omega^2 + \frac{15}{4}\beta\omega + 12b_1\omega = 0 \tag{16}$$

Which gives $a_1$ as

$$a_1 \approx -\frac{2\left(\pm\sqrt{450\beta^2 + 64(\omega^2 - 45\gamma)} + 8\omega\right)}{15\beta} \tag{17}$$

Rearranging Eq. (16) will give,

$$a_1^2 = \frac{8(-4a_1\omega + 15\beta + 48b_1)}{15\beta} \tag{18}$$

Also $N_1 = 0$ in Eq. (12) assuming $a_0 = 0$, up to $O(\varepsilon)$ gives,

$$-\frac{9}{8}a_1^2\beta - 12a_1\omega - b_1\omega^2 + 9\beta = 0 \tag{19}$$

Substituting (17) and (18) into (19) and rearranging gives the simplified bifurcation condition as:

$$\frac{32\omega\left(\sqrt{450\beta^2 + 64(\omega^2 - 45\gamma)} + 8\omega\right) + 10\gamma(5\omega^2 + 144)}{25\beta} = 0 \tag{20}$$

Importantly, equation (20) gives the conditions for stability transition of the symmetric backward solution. Bifurcation condition for different parameters is calculated from the approximate equation (20) and compared with numerical results in the following Fig. 3, appearing as dashed lines, showing excellent agreement with the numerical results. Numerical results are calculated from the original nonlinear system (3). We also calculated the exact bifurcation condition from Eq. (15), and its curves are not visually distinguishable in Fig. 3.

### 3.2.3 Asymptotic behavior of mean speed

In this section, the mean speed $V$ of the swimmer in the $\hat{\mathbf{x}}$ direction is calculated for the same asymptotic scaling approximation $\gamma \sim O(\varepsilon^2)$, $\beta \sim O(\varepsilon)$, and $\omega \sim O(\varepsilon^{-0.5})$ as in the previous section. The case shown below is for $(\bar{\theta}, \bar{\phi}) = (\pi, 0)$, and the same approach can be used for $(\bar{\theta}, \bar{\phi}) = (0,0)$. For calculating $V$, we need expressions for $\theta(t)$ and $\phi(t)$ in the symmetric periodic solution. For $\theta(t)$, using the same scaling as in (16) and expanding (8) up to 2nd order in $\theta$, and neglecting, retaining only terms up to order $O(\varepsilon)$, the symmetric periodic solution of (12) with $a_0 = 0$ is obtained as,

$$\begin{aligned}
a_1 &= \frac{2\left(\sqrt{2}\sqrt{9\beta^2(5\omega^2+144)^2 + 32\omega^2(\omega^2+144)^2} - 8\omega^3 - 1152\omega\right)}{3\beta(5\omega^2+144)} \\
b_1 &= \frac{32\omega\left(1152\omega + 8\omega^3 - \sqrt{2}\sqrt{32\omega^2(144+\omega^2)^2 + 9\beta^2(144+5\omega^2)^2}\right)}{\beta(144+5\omega^2)^2}
\end{aligned} \tag{21}$$

As for $\phi(t)$, again we seek harmonic balance solution as,

$$\phi(t) = c_1 \cos(t\omega) + d_1 \sin(t\omega) \tag{22}$$

the trigonometric functions of Eq. (5) in $\theta$ are expanded up to second order around $\theta = \pi$.

$$\ddot{\phi} + 6(\gamma(\pi - \theta) - 2\phi + \beta\left(-1 + \frac{1}{2}(\theta - \pi)^2\right)\sin(t\omega)) = 0 \tag{23}$$

Eq. (22) is substituted into (23), to solve $\phi(t)$ using harmonic balance (using the assumptions as in (10)) and the coefficients are obtained as follows.

$$c_1 = -\frac{3\left(-24a_1\beta b_1 + a_1^2\beta\omega - 96a_1\gamma + 3\beta b_1^2\omega + 8b_1\gamma\omega - 8\beta\omega\right)}{4\left(\omega^2 + 144\right)}$$

$$d_1 = \frac{3\left(a_1\beta b_1\omega + 6a_1^2\beta + 4a_1\gamma\omega + 18\beta b_1^2 + 48b_1\gamma - 48\beta\right)}{2\left(\omega^2 + 144\right)} \tag{24}$$

Then, it can be continued to find $\dot{x}(t)$ as a function of $a_1, b_1, c_1, d_1$. The mean speed $\dot{x}(t)$ in the $\hat{\mathbf{x}}$ direction can be obtained as (see Eq. (A13) in the Appendix):

$$\dot{x}(t) = -\frac{3\sin(\theta)\begin{pmatrix}\cos^2(\phi)(-\beta\cos(\theta)\sin(t\omega) + \gamma\sin(\theta) + \phi) - \\ -\left(\sin^2(\phi) + 5\right)(-\beta\cos(\theta)\sin(t\omega) + \gamma\sin(\theta) + \phi) + 4\phi\cos(\phi)\end{pmatrix}}{34 - 2\cos(2\phi)} -$$

$$-\frac{3\sin(\phi)\cos(\theta)(\cos(\phi)(-\beta\cos(\theta)\sin(t\omega) + \gamma\sin(\theta) + \phi) + 3\phi)}{\cos^2(\phi) - 9} \tag{25}$$

Now, expanding the trigonometric functions in (25) to first order about $\theta = \pi$ and $\phi = 0$ gives,

$$\dot{x}(t) = -\frac{3(\pi - \theta)\begin{pmatrix}(\beta\sin(t\omega) + \gamma(\pi - \theta) + \phi) - \\ -5(\beta\sin(t\omega) + \gamma(\pi - \theta) + \phi) + 4\phi\end{pmatrix}}{32} -$$

$$-\frac{3\phi\left(\left(\beta\sin(t\omega) + \gamma(\pi - \theta) + \phi\right) + 3\phi\right)}{8} \tag{26}$$

Substituting the $a_1, b_1$ series for $\theta(t)$ and the $c_1, d_1$ series for $\phi(t)$ from (10) and (22) into (26) and then rearranging and removing all oscillating terms to keep only the constant terms gives the final equation for $V$ as follows,

$$V \approx -\frac{3}{64}\left(4\beta b_1 + 16c_1^2 + 4\beta d_1 + 16d_1^2\right) \tag{27}$$

Substituting $a_1, b_1, c_1, d_1$ from (21) and (24) into (27), one obtains $V$ as a function of $\beta, \gamma$ and $\omega$. Eq. (27) gives the asymptotic approximation of $V$ for the backward solution with $(\bar{\theta}, \bar{\phi}) = (\pi, 0)$ and the expression for the forward solution $(\bar{\theta}, \bar{\phi}) = (0, 0)$, can be obtained in a similar way. The dashed curves in Fig. 5 (a) and (c) show the analytical prediction of $V$ for both forward (red) and backward (blue) solutions. It can be seen that there is good agreement between the analytic approximation and the numeric calculation of $V$ as a function of $\omega$ in Fig 5(a), which also qualitatively captures the optimum point with some deviation. In Fig. 5(c), for large $\beta$, the deviation between approximate and numeric calculation is very large since $\beta$ and $\gamma$ are large, and violate the scaling assumptions $\beta \sim O(\varepsilon)$ and $\gamma \ll \beta \ll 1 < \omega$ and it is not shown for higher values of $\beta$. The deviation of analytical calculation from numerical calculation is more evident in Fig 5(c), which plots $X$ as a function of $\beta$, where one can see a good agreement between approximate and numeric calculation only up to $\beta \sim 1.5$ for the backward motion, and larger deviation beyond this range, due to violation of the scaling assumptions.

## 4   Conclusion

The periodic dynamics of the microswimmer propulsion in the backward direction gives very interesting findings. In the forward direction, the motion is always stable, whereas in the backward direction ($\bar{\theta} = \pi$) the swimmer shows stability transition with subcritical pitchfork bifurcation, upon varying a single parameter out of $\beta, \gamma, \omega$ (see Fig. 4). Surprisingly, the swimmer can go faster in the backward direction than the forward direction and nonzero net propulsion exists for the case $\gamma = 0$. The parameter $\gamma$ can be tuned to obtain the optimum mean speed or displacement in the stable region, which calls for the scope of an experimental validation and gives hint towards its engineering applications in the future. Again, $\gamma$ is a very sensitive parameter in the system and the dynamics at $\gamma \to 0$ needs further investigation to get a full picture of the nonlinear dynamics in the domain.

The numerical approach successfully calculated the stability, bifurcation and optimum values of the swimmer's motion for the fixed point around $[\pi; 0]$, for different range of parameters. The harmonic balance approach together with small-angle approximation of $\phi$ as well as using scaling assumptions on $\beta, \gamma, \omega$, very well predicts the symmetric and asymmetric branches of the bifurcation, the mean speed of the swimmer with respect to the actuation frequency and also the bifurcation condition Eq. (20).

**Acknowledgment**

The research of YO is partially supported by Israel Ministry of Science and Technology, under grant no. 3-17383.

**Appendix A-Explicit calculation of dynamic equation of the two-link microswimmer.**

Here, we review the explicit derivation of the two-link swimmer's equations of motion, which were briefly formulated previously in [18,19]. By representing the forces and torques as, $\mathbf{F}_i = (\mathbf{f}_i, m_i)$ for $i = 1, 2$, Eq. (1) can be written in the matrix form as follows.

$$\mathbf{F}_i = -\mathbf{R}_i(\phi)\mathbf{V}_i;$$
$$\mathbf{V}_i = \begin{pmatrix} \mathbf{v}_i \\ \omega_i \end{pmatrix} = \mathbf{T}_i(\phi)\mathbf{v}_\mathbf{b} + \mathbf{E}_i(\phi)\dot{\phi} \tag{A1}$$

$\mathbf{v}_i$ is the vector of velocities, $\omega_i$ is the angular velocity and $\mathbf{v}_i, \mathbf{f}_i$ are expressed in the frame $(\hat{\mathbf{t}}, \hat{\mathbf{n}})$ attached to link-1, the head. Also, in (A1), $\mathbf{v}_\mathbf{b} = (v_t, v_n, \dot{\theta})^T$, where $v_t$ and $v_n$ are the tangential and normal velocities to the head and can be converted into the generalized coordinates by using the rotation matrix

$$\begin{pmatrix} \dot{x} \\ \dot{y} \end{pmatrix} = \begin{pmatrix} \cos\theta & \sin\theta & 0 \\ -\sin\theta & \cos\theta & 0 \\ 0 & 0 & 1 \end{pmatrix}^T \begin{pmatrix} v_t \\ v_n \end{pmatrix} \tag{A2}$$

In (A1), $\mathbf{R}_i$ represents the hydrodynamic resistance matrices and are given as [22],

$$\mathbf{R}_1(\phi) = c_t l \begin{bmatrix} 1 & 0 & 0 \\ 0 & 2 & 0 \\ 0 & 0 & \frac{1}{6}l^2 \end{bmatrix}; \mathbf{R}_2(\phi) = c_t l \begin{bmatrix} 1+\sin^2\phi & \cos\phi\sin\phi & 0 \\ \cos\phi\sin\phi & 1+\cos^2\phi & 0 \\ 0 & 0 & \frac{1}{6}l^2 \end{bmatrix} \quad (A3)$$

The terms $\mathbf{T}_i$ and $\mathbf{E}_i$ are described as following,

$$\mathbf{T}_1 = \begin{bmatrix} 1 & 0 & 0 \\ 0 & 1 & 0 \\ 0 & 0 & 1 \end{bmatrix}; \mathbf{E}_1 = \begin{bmatrix} 0 \\ 0 \\ 0 \end{bmatrix}$$

$$\mathbf{T}_2 = \begin{bmatrix} 1 & 0 & -0.5l\sin\phi \\ 0 & 1 & -0.5l - 0.5l\cos\phi \\ 0 & 0 & 1 \end{bmatrix}; \mathbf{E}_2 = \begin{bmatrix} 0.5l\sin\phi \\ 0.5l\cos\phi \\ -1 \end{bmatrix} \quad (A4)$$

The net hydrodynamic force on the i[th] link can be represented as,

$$\mathbf{F}_{hyd,i} = -\mathbf{R}_i\left(\mathbf{T}_i\mathbf{v_b} + \mathbf{E}_i\dot{\phi}\right) \quad (A5)$$

Similarly, the generalized magnetic force vector acting on the i[th] link and the torque acting at the joint are given by,

$$\mathbf{F}_b = \begin{bmatrix} 0 \\ 0 \\ \mathbf{L}\cdot\hat{\mathbf{z}} \end{bmatrix}; \tau = -k\phi \quad (A6)$$

Again, assuming the swimmer moves quasistatically, the net forces and torques on each link will be zero.

$$\sum \mathbf{T}_i^T\left(\mathbf{F}_{hyd,i} + \mathbf{F}_{b,i}\right) = 0; \sum \mathbf{E}_i^T\left(\mathbf{F}_{hyd,i} + \mathbf{F}_{b,i}\right) + \tau = 0 \quad (A7)$$

Substituting (A1), (A3), (A4) and (A6) in (A7) gives,

$$\mathbf{R}_{bb}\mathbf{v_b} + \mathbf{R}_{bu}\dot{\phi} = \sum \mathbf{T}_i^T\mathbf{F}_b; \mathbf{R}_{bu}\mathbf{v_b} + \mathbf{R}_{uu}\dot{\phi} = \sum E_i^T\mathbf{F}_b - k\phi \quad (A8)$$

where $\mathbf{R}_{bb} = \sum \mathbf{T}_i^T\mathbf{R}_i\mathbf{T}_i; \mathbf{R}_{bu} = \sum \mathbf{T}_i^T\mathbf{R}_i\mathbf{E}_i; \mathbf{R}_{uu} = \sum \mathbf{E}_i^T\mathbf{R}_i\mathbf{E}_i$

We can write the above equations in a concise matrix form as follows,

$$\mathbf{A}\begin{pmatrix}\mathbf{v_b}\\\dot{\phi}\end{pmatrix}=\mathbf{b};\mathbf{A}=\begin{pmatrix}\mathbf{R}_{bb}&\mathbf{R}_{bu}\\\mathbf{R}_{bu}^{T}&\mathbf{R}_{uu}\end{pmatrix};\mathbf{b}=\begin{pmatrix}\sum\mathbf{T}_i^T\mathbf{F}_{b,i}\\\sum\mathbf{E}_i^T\mathbf{F}_{b,i}-k\phi\end{pmatrix} \quad (A9)$$

$$\mathbf{A}=c_t l\begin{pmatrix}\sin^2(\phi)+2 & \sin(\phi)\cos(\phi) & -\dfrac{l}{2}\sin(\phi)(\cos(\phi)+2) & l\sin(\phi)\\[4pt]\sin(\phi)\cos(\phi) & \cos^2(\phi)+3 & -2l\cos^4\left(\dfrac{\phi}{2}\right) & l\cos(\phi)\\[4pt]-\dfrac{l}{2}\sin(\phi)(\cos(\phi)+2) & -2l\cos^4\left(\dfrac{\phi}{2}\right) & \dfrac{l^2}{24}(24\cos(\phi)+3\cos(2\phi)+29) & \dfrac{l^2}{6}(-3\cos(\phi)-4)\\[4pt]l\sin(\phi) & l\cos(\phi) & \dfrac{l^2}{6}(-3\cos(\phi)-4) & \dfrac{2l^2}{3}\end{pmatrix} \quad (A10)$$

$$\mathbf{b}=\begin{pmatrix}0\\0\\h(b\cos(\theta)\sin(t\Omega)-c\sin(\theta))\\-k\phi\end{pmatrix} \quad (A11)$$

$$\begin{pmatrix}v_t\\v_n\\\dot{\theta}\\\dot{\phi}\end{pmatrix}=\begin{pmatrix}-\dfrac{3\sin(\phi)(\cos(\phi)(-bh\cos(\theta)\sin(t\Omega)+ch\sin(\theta)+k\phi)+3k\phi)}{l^2 c_t(\cos(\phi)-3)(\cos(\phi)+3)}\\[6pt]-\dfrac{3\left(\cos^2(\phi)(-bh\cos(\theta)\sin(t\Omega)+ch\sin(\theta)+k\phi)-\left(\sin^2(\phi)+5\right)(-bh\cos(\theta)\sin(t\Omega)+ch\sin(\theta)+k\phi)+4k\phi\cos(\phi)\right)}{2l^2 c_t(\cos(2\phi)-17)}\\[6pt]\dfrac{3\cos^2(\phi)(-bh\cos(\theta)\sin(t\Omega)+ch\sin(\theta)+k\phi)-3\left(\sin^2(\phi)-19\right)(-bh\cos(\theta)\sin(t\Omega)+ch\sin(\theta)+k\phi)+36k\phi\cos(\phi)}{l^3 c_t(\cos(2\phi)-17)}\\[6pt]\dfrac{6(\cos(\phi)+3)^2(-bh\cos(\theta)\sin(t\Omega)+ch\sin(\theta)+2k\phi)}{l^3 c_t(\cos(2\phi)-17)}\end{pmatrix} \quad (A12)$$

Using the relation (A2), we obtain:

$$\begin{pmatrix}\dot{x}\\\dot{y}\end{pmatrix}=\dfrac{1}{c_t l^2}\begin{pmatrix}\dfrac{3\sin(\theta)\left(\begin{array}{c}\cos^2(\phi)(-bh\cos(\theta)\sin(t\Omega)+ch\sin(\theta)+k\phi)-\\-\left(\sin^2(\phi)+5\right)(-bh\cos(\theta)\sin(t\Omega)+ch\sin(\theta)+k\phi)+4k\phi\cos(\phi)\end{array}\right)}{2(\cos(2\phi)-17)}-\\[6pt]-\dfrac{3\cos(\theta)\sin(\phi)(\cos(\phi)(-bh\cos(\theta)\sin(t\Omega)+ch\sin(\theta)+k\phi)+3k\phi)}{(\cos(\phi)-3)(\cos(\phi)+3)}\\[6pt]-\dfrac{3\cos(\theta)\left(\begin{array}{c}\cos^2(\phi)(-bh\cos(\theta)\sin(t\Omega)+ch\sin(\theta)+k\phi)-\\-\left(\sin^2(\phi)+5\right)(-bh\cos(\theta)\sin(t\Omega)+ch\sin(\theta)+k\phi)+4k\phi\cos(\phi)\end{array}\right)}{2(\cos(2\phi)-17)}-\\[6pt]-\dfrac{3\sin(\theta)\sin(\phi)(\cos(\phi)(-bh\cos(\theta)\sin(t\Omega)+ch\sin(\theta)+k\phi)+3k\phi)}{(\cos(\phi)-3)(\cos(\phi)+3)}\end{pmatrix} \quad (A13)$$

Using the nondimensionalization introduced in Section 2, one obtains Eq. (25).